\documentclass[aps,eqsecnum,apsi,preprintnumbers,showpacs]{revtex4}
\usepackage{graphicx}
\usepackage{latexsym}
\usepackage{amsmath}
\usepackage{amssymb}
\usepackage{epsfig}
\usepackage{longtable}

\newcommand{\be}{\begin{equation}}
\newcommand{\ee}{\end{equation}}
\newcommand{\bea}{\begin{eqnarray}}
\newcommand{\eea}{\end{eqnarray}}

\def\tev{\, \, {\rm TeV}}
\def\gev{\, \, {\rm GeV}}

\begin{document}

\preprint {RECAPP-HRI-2011-001}

\title{\Large \bf Low-scale SUSY breaking by modular fields and Higgs mass bounds}

\author{Satyanarayan Mukhopadhyay$^{a}$ \footnote{E-mail:
    satya@hri.res.in}, Biswarup Mukhopadhyaya$^{a}$ \footnote{E-mail:
    biswarup@hri.res.in}, Soumitra SenGupta$^{b}$ \footnote{E-mail:
    tpssg@iacs.res.in} } \affiliation{$^a$Regional Centre for
  Accelerator-based Particle Physics, Harish-Chandra Research
  Institute, Chhatnag Road, Jhusi, Allahabad - 211 019,
  India.\\ $^b$Department of Theoretical Physics,\\Indian Association
  for the Cultivation of Science, Kolkata-700 032, India. }

\begin{abstract}
We consider a scenario where supersymmetry (SUSY) is broken at a
relatively low scale by modular fields of extra compact spacelike
dimensions. The effect of both soft and hard SUSY breaking terms on
the mass of the lightest neutral Higgs boson are investigated.  An
important conclusion is that the lightest neutral Higgs can be
considerably more massive than what is expected in the MSSM, if the
overseeing theory breaks SUSY at a scale not too far above a TeV. An
explicit model that implements this has been shown for illustration.
\end{abstract}

\maketitle

\section{Introduction}
\label{intro}

Supersymmetry (SUSY), as a cure to the quadratic divergence problem,
requires embedding in a bigger canvas, in order to account for the
many free parameters that appear in the soft SUSY-breaking
Lagrangian. Also, such embedding enables one to generate
phenomenologically consistent new particle spectra, bypassing the
supertrace theorem.  Some common schemes that are frequently
considered include high scale SUSY breaking through supergravity
(SUGRA), or SUSY breaking via gauge mediation (GMSB) at a somewhat
lower scale, with the standard model (SM) gauge interactions yielding
SUSY-breaking masses via loop-induced interaction with a messenger
sector.

In order to be phenomenologically relevant, the new particles in the
minimal SUSY standard model (MSSM) should have masses around the TeV
scale.  In spite of the exalted `top-down' approach, it is somewhat
disquieting that such a striking feature of nature as boson-fermion
symmetry (as well as its controlled breaking) should be determined
entirely by scales several orders of magnitude higher.  Historically,
as we have been able to access progressively higher energy scales, we
have seen new laws of physics gradually unravelled. Furthermore, the
physics of a particular scale is found to be mostly influenced by that
in the scales immediately above it.  The question is: when it comes to
the outstanding questions of SUSY phenomenology at the TeV scale,
wouldn't it be fair to expect the decisive clues to lie around 10 TeV
or so?

Let us look, for the sake of illustration, at two issues related to
MSSM, following such a bottom-up philosophy.  First, the
SUSY-conserving Higgsino mass parameter $\mu$ occurring in the
superpotential should naturally be very large if the MSSM is embedded
in a theory valid upto a very large (Planck or Grand Unification)
scale. The fact that phenomenology demands a $\mu$ within a TeV leads
to a new naturalness problem.  This `$\mu$-problem' however, is not so
acute anymore if MSSM emerges with SUSY breaking terms given by some
effective theory, valid upto about 10 TeV.  Secondly, according to
accepted dicta, the non-observation of the lightest neutral Higgs upto
about 140 GeV can almost rule out MSSM, for radiative corrections
within the model cannot hike it beyond such a value.  However, one may
enquire whether some additional physics around 10 TeV can boost the
lightest neutral Higgs mass to still higher values.  It has indeed
been claimed in an effective theory framework that higher-order terms
in the MSSM fields, suppressed by a scale somewhat above a TeV,
provide some upward revisions to the Higgs
mass~\cite{antoniadis,seiberg,casas,casas-1}.  There have also been
several studies in recent times, suggesting modifications to Higgs
mass upper bounds when the standard model gauge group is extended by,
for example, additional U(1) factors~\cite{U_1}.

It should be noted that the potential threat from flavour changing
neutral currents (FCNC) in SUGRA-type scenarios can be avoided if
flavour diagonal soft SUSY breaking masses are generated at a
relatively low scale. This is because the mass parameters then do not
`run' long enough to cause sufficient misalignment between the fermion
and sfermion mass matrices.

The other point to remember is that one can in principle also allow
hard SUSY-breaking terms in the Lagrangian~\cite{Martin}. Such terms
may acquire particular significance when one has additional physics
intervening at relatively low scale(s). The role of such terms in the
context of neutrino masses~\cite{neutrino} and also in other contexts
of SUSY phenomenology~\cite{haber-HSB} have been investigated in the
literature. It has been shown that, with specific F-term SUSY breaking
assumptions, the hard terms can cause radiative corrections to the
Higgs mass which are at best of the magnitude of the soft SUSY
breaking masses themselves~\cite{Polonsky_1}. In addition, as the
scale of SUSY breaking is now low, these terms can give rise to
substantial tree-level corrections to the MSSM Higgs quartic
couplings~\cite{casas,Polonsky_1}. In a completely model-independent
setting, however, the radiative corrections arising due to the
presence of hard terms can be unacceptably large. On the other hand,
if MSSM is embedded in an effective theory valid upto about 10 TeV,
then hard SUSY breaking terms that are non-vanishing below such a
scale are always safe from the viewpoint of the lightest neutral Higgs
mass, irrespective of whether they are arising from F-term vacuum
expectation values (vev) or not. They can, at the same time, be
instrumental in contributing substantially and positively to the Higgs
mass. If that happens, then the non-observation of the lightest
neutral Higgs within the stipulated mass limit is possible, though one
may see other signals of new physics at the Large Hadron Collider.  We
discuss such a possibility in this paper, in the context of a scenario
based on extra compact spacelike dimensions in a string inspired
higher dimensional supergravity model. As will be seen below, the
essence of our proposal lies in treating the vev's of the modular
fields for the extra dimensions as the cut-off of the low energy
theory below TeV scale and therefore as a suppressant in the terms in
the Kahler potential and the superpotential leading to soft SUSY
breaking.

In any theory of this type, the pertinent question is: what provides
the scale of 10 TeV in a sector beyond the MSSM spectrum? In the extra
dimensional context, one usually thinks of the `bulk' Planck scale
which can be low.  We consider an alternative possibility
here. Whenever the additional spacelike dimensions are compactified,
one ends up with modular fields whose vev's are related to the stable
radii of these dimensions.  There can be a number of modular fields,
including the radii of the different compact dimensions and the angles
among them. We suggest the possibility of the vev's of scalar
components of the modular superfields as the additional scale.  If
these vev's are about 10 TeV, then the SUSY theory is effective below
this scale, above which the modular fields develop as dynamical
degrees of freedom. Below the energy scale corresponding to the vev of
the modular fields, on the other hand, one can think of an effective
description, with all higher-dimensional operators suppressed by the
aforementioned scale.  The dominant contributions to soft SUSY
breaking terms in the observable sector may arise from such terms
which take the place of terms suppressed by the Planck scale in common
supergravity scenarios.

Here we suggest a KKLT-type~\cite{kklt} scenario where SUSY is broken
by introducing a lifting term in the potential in terms of the modulus
field $T$ which lifts an N=1 SUSY anti-de Sitter vacuum to a
Minkowski/de-Sitter vacuum with broken SUSY.  The corresponding scalar
potential has a well-defined minimum and the vev of the $T$ field can
be tuned at a scale near $10 \tev$.  The relevant F-term component is
at an intermediate scale of around $\sim (3 \tev)^2$ and one can
easily obtain the required low-energy SUSY spectrum with the soft
masses $\sim 1 \tev$. As has been already mentioned, if SUSY is broken
at a relatively low scale, the effect of certain hard SUSY-breaking
operators are no longer negligible, although they do not bring back
any harmful quadratic divergences and can give finite and bounded
contributions to observables like the Higgs mass. In order to see the
implications of such a low-scale SUSY-breaking scenario, we shall
study how the upper bounds on the lightest Higgs boson mass might be
modified, thereby reducing the tension between this bound in MSSM and
the limit coming from LEP.

The framework proposed by us is outlined more precisely in
section~\ref{general}.  In section~\ref{model}, a form of the
potential that can lead to the appropriate vev's of the scalar and
auxiliary components of the modular fields is suggested.  The
implications on the lightest neutral Higgs mass, including the effects
of the hard terms, are shown in section~\ref{higgs-mass}, along with
some numerical estimates.  We summarise and conclude in section
~\ref{summary}.

\section{The general scenario}
\label{general}

In a generic SUSY breaking mechanism two different scales are
involved. One is the SUSY breaking scale $\sqrt{F}$ which corresponds
to the vev's of the relevant auxiliary fields in the SUSY breaking
sector. The other one is the $M$, associated with the interactions
that transmit the breaking to the observable sector. In an extra
dimensional model $M$ is determined by the compactification scale
which in turn is related to the inverse of the radius of the extra
dimension. $M$ thus sets the scale of the effective theory and as we
are considering an effective theory below $M$, this transmission takes
place through higher-dimensional operators suppressed by powers of
$M$.  For example, these operators give rise to the scalar soft masses
of the form
\begin{equation}
 m_{soft} \sim \frac{F}{M}.
\end{equation}
 In addition to the soft terms, generically hard SUSY breaking terms
 can also be present in the effective theory. For example, one can
 have a scalar quartic coupling of the form
\begin{equation}
 \lambda_{hard} \sim \frac{F^2}{M^4} \sim \frac{m^2_{soft}}{M^2}
\end{equation}
Note that the quadratic correction due to this term to the lightest
Higgs boson mass is ${\cal{O}}(m^2_{soft})$, and thus the mass shift
is not inordinately high even if one has large $M$. A similar
observation can be made even if one has

\begin{equation}
 \lambda_{hard} \sim \frac{F}{M^2},
\end{equation}

as the net loop-induced contribution in a gauge-invariant framework
finally turns out to be ${\cal{O}}(m^2_{soft})$.  This can be
understood from the fact that the SUSY-breaking operators with such
couplings give only holomorphic corrections to the MSSM Higgs
potential~\cite{seiberg}.  However, if one turns completely model
independent, and assumes that a quartic SUSY breaking term
proportional to a purely phenomenological parameter $\lambda$ is
induced in the effective theory below the scale $M$, then the shift in
the lightest Higgs mass proportional to $M^2$ cannot in general be
avoided.  Thus, the reconciliation between phenomenological hard SUSY
breaking terms and a manageable shift in the Higgs mass is best
achieved if the scale relevant for SUSY breaking is within an order
above a TeV.  At the same time, the $\mu$ parameter, which needs to be
at best about a TeV, no more raises a naturalness issue.

There are two relevant scales determining the SUSY-breaking
parameters, namely, $\sqrt{F}$ and $M$. The only phenomenological
input that we have is that $m_{soft} = \mathcal{O} (1\tev)$, which is
required if TeV-scale SUSY is to be a solution to the naturalness
problem. As has been observed before in models of low-scale SUSY
breaking, the MSSM assumption of a hierarchy of scales is not really
necessary and therefore the hard terms can also be
non-negligible. Thus for example, even at the tree level, the Higgs
quartic coupling can get enhanced and the theoretical upper bound on
the lightest Higgs mass modified accordingly.

Here we consider an extra-dimensional scenario in which both the
scales $\sqrt{F}$ and $M$ are of similar order, on the order of
several $\tev$'s.  In studies of this kind, the `bulk' Planck mass is
hypothesised as the source of the scale $M$.  As has been already
mentioned, we seek an alternative scenario where gravity does not have
the primary role; instead, it is the modular fields which guide us to
the scale of the effective theory related to the radius of
compactification.

In the framework considered here, the universe is $(4+n)$ dimensional
with $n\geq 1$ extra compact spacelike dimensions. While gravity can
propagate in the new dimensions, the standard model (SM) fields are
assumed to be localized on a 3-brane in the higher-dimensional
space. The radii of these compact dimensions act like modular fields
($T_i$) and the stable radii of these extra dimensions are related to
vev's $\langle {T_i}\rangle$ of these modular fields. As the compact
dimensions are considered to be large, generically $\langle T_i\rangle
<< M_{Pl}$, where $M_{Pl}$ is the four-dimensional Planck scale. Now,
SUSY is assumed to be broken by the vev's of the F-term components of
the fields $T_i$ , $\langle {F_{T_i}}\rangle$. The scale of the
effective 4-d theory in the visible sector however is set by $\langle
{T_i}\rangle$. With both $\langle{T_i}\rangle$ and $\langle
{F_{T_i}}\rangle$ an order or two above the TeV scale, the
phenomenology of the `effective' MSSM is controlled by a low-lying
scale, with all the merits that have been mentioned above.

In order to have $m_{soft} = \mathcal{O} (1\tev)$, where the mediation
scale $M \sim 10 \tev$, we require $F \sim 10 \tev^2$, i.e., $\sqrt{F}
\sim 3 \tev$. This is a constraint that our model has to satisfy. For
simplicity we assume a common stable radius for all the extra
dimensions and therefore a common vev for all the moduli
fields. Therefore, $\sqrt{\langle {F_T} \rangle}$, has to be of the
same order as $\langle T\rangle$. In section~\ref{model} we shall
construct an illustrative model and show that such a thing is easily
achievable.

It should also be mentioned that such a scenario has a testable
difference from one where the low-valued bulk Planck mass (such as in
the scenario proposed by Antoniadis, Arkani-Hamed, Dimopoulos and
Dvali~\cite{ADD}) is the scale of the effective theory.  In the latter
situation, the convolution of any amplitude involving the emission of
gravitons by the density of graviton states results in the total
amplitude for graviton emission suppressed by the bulk Planck scale
only, thus raising hopes for the signatures of gravitons at the
LHC. In the situation we consider, the bulk Planck mass is
considerably higher.  As a result, missing energy signals involving
gravitons are not going to be visible, although all SUSY signals
appropriate for the mass spectrum are predicted as usual.

Before we go into the Higgs spectrum, we outline some features of our
proposal in the next section. Some related approaches, also based on
SUSY breaking via extra dimensions, can be found
in~\cite{Extrad-SUSY}.  There have also been studies on the
modification of lightest Higgs boson mass in the context of SUSY
theories formulated in higher space dimensions~\cite{Extrad-higgs}.
%%%%%%%%%%%%%%%%%%%%%%%%%%%%%%%%%%%%%%%%%%%%%%%%%%%%%%%%%%%%%%%%%%%%%%%%%%%%%%%%%%%%%%%%%%%%%%%%%%%%%%%%%%%

\section{A specific model}
\label{model}

%%%%%%%%%%%%%%%%%%%%%%%%%%%%%%%%%%%%%%%%%%%%%%%%%%%%%%%%%%%%%%%%%%%%%%%%%%%%%%%%%%%%%%%%%%%%%%%%%%%%%%%%%%%
%%%%%%%%%%%%%%%%%%%%%%%%%%%%%%%%%%%%%%%%%%%%%%%%%%%%%%%%%%%%%%%%%%%%%%%%%%%%%%%%%%%%%%%%%%%%%%%%%%%%%%%%%%%
We now illustrate our analysis with a model which can be motivated in
an underlying string-inspired supergravity theory~\cite{choi}.

The model has a KKLT-like~\cite{kklt} setup where light modulus,
namely the volume modulus $T$, is stabilized by non-perturbative
effects like gaugino condensation~\cite{ibanez} leading to an $N=1$
supersymmetric anti-de-Sitter (AdS) vacuum.  This AdS vacuum is then
uplifted to a SUSY breaking Minkowski (or de-Sitter) vacuum by branes
which break $N=1$ SUSY explicitly~\cite{klebanov}. The vev of the
modulus $T$ as well as the corresponding $F$-term can be set to
desired values, which in turn can generate soft terms in the visible
sector at the TeV scale.

In our model, the effective $N=1$ SUGRA contains the $T$-modulus, and
the effective description breaks down at the compactification scale
$M_c$, which, for a large radius (or light moduli) compactification
may be set to be about $10 \tev$.

Following KKLT proposal we introduce a SUSY breaking $\overline{D3}$
brane which uplifts the AdS vacuum to a dS vacuum where the vev of the
$T$ field (related to $M_c$) is primarily determined by the $N=1$ SUSY
sector.  In our model, the other modulus field like the dilaton is
assumed to have a large vev and hence it decouples from the theory
below the scale $\langle T \rangle$.

The $T$-sector has the Kahler potential $K$ and the superpotential $W$
given by

\begin{equation}
 K=-3\ln (T+\overline{T})
\end{equation}
and

\begin{equation}
W=W_0 - Ae^{-aT},
\end{equation}

where $T=t+i\tau$, and T, the Kahler potential and the superpotential
have been appropriately scaled to make them dimensionless. Here $\tau$
is the axion present in the theory.  Since the Kahler potential
depends only on $t$, therefore, the overall phase of $W$ is
irrelevant. Moreover, the relative phase between $W_0$ and $A$ can be
eliminated by shifting the axion field $\tau$ such that

\begin{equation}
 \langle \tau \rangle =0. 
\end{equation}

The condition for unbroken SUSY is given by,
\begin{equation}
\label{unbroken_susy}
 \langle D_T W \rangle =0,
\end{equation}
where
\begin{equation}
 D_T W = {\partial_T} W + \frac{\partial K}{\partial T} W.
\end{equation}

Solving for $W_0$, we get
\begin{equation}
\label{W0}
 W_0 = \langle A e^{-at} (1+\frac{2at}{3}) \rangle.
\end{equation}

Using eqn.~\ref{W0}, we can write the following approximate relation
determining $\langle t \rangle$ (which is equal to $\langle T
\rangle$, as the axion vev is zero in this case) in terms of the
parameters $a$, $A$ and $W_0$:
\begin{equation}
\label{at}
 \langle at \rangle \simeq \ln \frac{A}{W_0}.
\end{equation}

The scalar potential for the $T$ field can be calculated using the
following expression
\begin{equation}
 V= M_{Pl}^4 e^{K} [K^{T \overline{T}} |D_T W|^2 - 3 |W|^2],
\end{equation}

where $K^{T \overline{T}} = (\frac{\partial^2 K}{\partial \overline{T}
  \partial T})^{-1}$.  It can be shown that, at the SUSY preserving
vacuum, defined by eqn.~\ref{unbroken_susy}, the minimum of the
potential takes the value
\begin{equation}
 \langle V \rangle = -3 m^2_{3/2} M^2_{Pl},
\end{equation}
  
which is an AdS vacuum. Here, $m_{3/2}$ is the gravitino mass. 

To stabilize $T$, we now introduce a $\overline{D3}$ brane which gives
rise to a lifting term $D/{t^n}$ in the scalar potential. For small
values of $t$, this gives a large contribution to the potential.

SUSY is now broken in this hidden sector with an $F$-term vev given by
\begin{equation}
\label{F_T}
 F_T \simeq \frac{n}{a} m_{3/2}.
\end{equation}

In order to obtain soft SUSY breaking masses of the order of $1 \tev$,
we need to impose the condition that $m_{soft} \sim F_T/\langle T
\rangle \sim 1 \tev$. If, in addition, we take $m_{3/2} \sim 1 \tev$,
we see from eqns.~\ref{at} and ~\ref{F_T} that for $n=2$, we shall
need $W_0$ and $A$ to satisfy the relation
\begin{equation}
 \ln \frac{A}{W_0} \sim 2.
\end{equation}
 
Such values of $W_0$ and $A$ are easily achievable in the framework we
consider. Also note that, by making a suitable choice of the parameter
$a$, one can also obtain the desired values for $F_T$ and $\langle T
\rangle$. For example, with $a=0.2 \tev^{-1}$, we can get $\langle T
\rangle = 10 \tev$, and correspondingly, $F_T = 10 \tev^2$, as
required.

Thus, we find by constructing this illustrative model that one can
obtain a suitable low-scale SUSY breaking framework where the vev of
the modular fields can act as the effective scale of suppression for
the soft and hard SUSY breaking operators.  Similar conclusion can
also be obtained in other scenarios where the dilaton sector also
acquires a superpotential from gaugino condensation.

%%%%%%%%%%%%%%%%%%%%%%%%%%%%%%%%%%%%%%%%%%%%%%%%%%%%%%%%%%%%%%%%%%%%%%%%%%%%%%%%%%%%%%%%%%%%%%%%%%%%%%%%%%%%
%%%%%%%%%%%%%%%%%%%%%%%%%%%%%%%%%%%%%%%%%%%%%%%%%%%%%%%%%%%%%%%%%%%%%%%%%%%%%%%%%%%%%%%%%%%%%%%%%%%%%%
%%%%%%%%%%%%%%%%%%%%%%%%%%%%%%%%%%%%%%%%%%%%%%%%%%%%%%%%%%%%%%%%%%%%%%%%%%%%%%%%%%%%%%%%%%%%%%%%%%%%%%%%%%%
\section{Upper bound on the lightest neutral Higgs mass}
\label{higgs-mass}

We have outlined above a scenario where the vev of the $T$ fields sets
the scale of suppression for nonrenormalisable terms responsible for
soft as well as hard SUSY beaking in the visible sector. Before we go
on to examine the modified upper bounds on the lightest neutral Higgs
mass in such a scenario, let us recall the salient features of the
MSSM Higgs sector~\cite{Djouadi_2}. One requires two $SU(2)_L$ Higgs
doublets, $H_2$ and $H_1$, with hypercharge $\pm 1$. The scalar
potential receives contributions from the F-terms, D-terms and the
soft SUSY breaking terms. The part of the tree-level potential
exclusive to the electrically neutral components of the Higgs fields
$H^0_{1,2}$ is given by
\begin{equation}
\label{MSSM_V_0}
 V_{MSSM}^{H^0}=\overline{m_1}^2|H_1^0|^2+\overline{m_2}^2|H_2^0|^2-\overline{m_3}^2(H_1^0H_2^0+h.c.)+
 \frac{1}{8}(g_1^2+g_2^2)(|H_1^0|^2-|H_2^0|^2)^2
\end{equation}
with $\overline{m}_{1,2}^2=|\mu|^2+m^2_{H_{1,2}}$ and
$\overline{m}_{3}^2=B\mu$ where $m^2_{H_{1,2}}$ and $B$ are soft SUSY
breaking parameters and $\mu$ is the Higgsino mass term in the
superpotential. Also, $g_1$ and $g_2$ are the $U(1)_Y$ and $SU(2)_L$
gauge couplings respectively.

It has been mentioned that, in addition to the soft SUSY breaking
contributions included in the above potential, one can in principle
have additional hard SUSY breaking terms as well. This results in
additional quartic terms in the superpotential, whose coefficients are
not related to the gauge couplings. Such dimensionless SUSY breaking
couplings do not, however, give rise to inordinately high quadratic
corrections to the Higgs mass, so long as they are proportional to
powers of $F/M$, although they can have more catastrophic consequences
in the most general situation. These couplings can become particularly
important in models of low-scale SUSY breaking where they are not
suppressed by a very high-scale.

All possible renormalizable supersymmetry breaking interactions have
been classified, for example, in~\cite{Martin}. We concentrate on the
hard SUSY breaking terms arising only in the Higgs sector of the
Lagrangian. This is partly for the sake of simplification, and partly
due to the fact that the modified bounds on the Higgs mass suggested
by us depend on other hard SUSY breaking terms only at higher orders
of perturbation.

Following ref.~\cite{Martin}, we consider the possible hard SUSY
breaking interactions involving the Higgs scalar fields. In the SUSY
Higgs sector, the possible gauge-invariant terms are given by

\begin{equation}
\label{L_hard}
 -\mathcal{L}_{Hard}=\frac{F}{M^2}[(H_2.H_1)^2+h.c.]+\frac{|F|^2}{M^4} 
[(H_1^{\dagger}H_1)^2+(H_2^{\dagger}H_2)^2+(H_1^{\dagger}H_1)(H_2^{\dagger}H_2)+
(H_1^{\dagger}H_2)(H_2^{\dagger}H_1)]
\end{equation}
  
In the term proportional to $\frac{|F|^2}{M^4}$, we see that all
possible quartic terms involving the Higgs fields in the MSSM scalar
potential can now arise via the dimensionless SUSY breaking couplings
also. Note that, in the MSSM scalar potential, these terms arise from
the D-term contributions of the $SU(2)_L$ and $U(1)_Y$
interactions. Thus the coefficients of the quartic terms are
determined entirely in terms of the gauge couplings $g_1$ and $g_2$
(see eqn.~\ref{MSSM_V_0}). In contrast, the above SUSY breaking
quartic terms reintroduce one-loop contributions to the Higgs
mass(es), which can potentially be quadratically divergent.

Due to the addition of these new hard SUSY breaking terms given in
eqn.~\ref{L_hard}, the tree-level MSSM scalar potential involving the
neutral components of the Higgs doublets gets modified. The modified
potential, $V^{H_0}$, can now be written as

\begin{equation}
 V^{H^0}=V_{MSSM}^{H^0}+V_{Hard}^{H^0},
\end{equation}
 where 
\begin{equation}
\label{V_Hard}
 V_{Hard}^{H^0}=\epsilon_1 [(H_1^0 H_2^0)^2 +h.c.] + \epsilon_2
 (|H_1^0|^4+|H_2^0|^4+|H_1^0|^2|H_2^0|^2).
\end{equation}

Here we have defined

\begin{eqnarray}
 \epsilon_1 &=& \frac{F}{M^2} \nonumber \\
\epsilon_2  &=& \frac{|F|^2}{M^4}.
\end{eqnarray}

We now assume that at the minimum of the potential $V^{H^0}$ the
neutral components of the two Higgs fields develop vacuum expectation
values
\begin{equation}
 \langle H_1^0\rangle=\frac{v_1}{\sqrt{2}}, ~~~~~~~\langle
 H_2^0\rangle=\frac{v_2}{\sqrt{2}}.
\end{equation}

In order to trigger electroweak symmetry breaking at the right scale,
$v_1$ and $v_2$ must satisfy the relation
\begin{equation}
 (v_1^2+v_2^2)=v^2=\frac{4M_Z^2}{g_1^2+g_2^2}=(246 \gev)^2.
\end{equation}

Given the parameters in the soft and hard SUSY breaking sectors, $v_1$
and $v_2$ can be determined by using the potential minimization
conditions:
\begin{equation}
 \frac{\partial V^{H^0}}{\partial H_1^0}=\frac{\partial
   V^{H^0}}{\partial H_2^0}=0,
\end{equation}
which in this case translate to:
\begin{eqnarray}
\label{minimize}
 \overline{m_1}^2-\overline{m_3}^2 \tan \beta + \frac{M_Z^2}{2}\cos 2 \beta +
 \epsilon v^2 &=& 0,\\ \overline{m_2}^2-\overline{m_3}^2 \cot \beta -
 \frac{M_Z^2}{2}\cos 2 \beta + \epsilon^{\prime} v^2 &=& 0,
\end{eqnarray}

where
\begin{eqnarray}
\tan\beta &=& \frac{v_2}{v_1},\\ \epsilon &=&
(\epsilon_1+\frac{\epsilon_2}{2})\sin^2\beta + \epsilon_2 \cos^2
\beta\\ \epsilon^{\prime} &=&
(\epsilon_1+\frac{\epsilon_2}{2})\cos^2\beta + \epsilon_2 \sin^2 \beta
\end{eqnarray}

All the above relations have been written with the assumption that
there is no CP-violation in the Higgs sector at the tree
level. Therefore, all the relevant parameters and in particular, the
vev's of the Higgs fields can be chosen as real.

The mass matrix for the Higgs bosons $\mathcal{M}^2_{ij}$ can be
computed using the following relation
\begin{equation}
\label{mass_1}
 \mathcal{M}^2_{ij}=\frac{\partial^2V^{H^0}}{\partial H^0_i H^0_j}
 \mid_{\langle H_1^0\rangle=\frac{v_1}{\sqrt{2}},\langle
   H_2^0\rangle=\frac{v_2}{\sqrt{2}}}.
\end{equation}

Following the usual convention in MSSM, we express the Higgs mass
eigenvalues in terms of two free parameters in the Higgs sector, which
we take to be $M_A$ and $\tan \beta$. We then consider the bound on
the Higgs mass in the so-called decoupling limit where $M_A>>M_Z$, and
the expression for the lightest neutral Higgs mass has a rather simple
form. Later in this section, where we present numerical results in
Table~\ref{tab1}, we show the Higgs mass shifts, based on an exact
numerical calculation, away from the decoupling limit. We now need to
calculate the pseudoscalar Higgs mass $M_A$. The neutral pseudoscalar
mass matrix is given by

\begin{equation}
\mathcal{M}^2_{Im}=\left( \begin{array}{cc}
  \overline{m_3}^2\tan\beta-2\epsilon_1 v^2 \sin^2\beta &
  \overline{m_3}^2-2\epsilon_1 v^2 \sin \beta \cos \beta
  \\ \overline{m_3}^2-2\epsilon_1 v^2 \sin \beta \cos \beta &
  \overline{m_3}^2\cot\beta-2\epsilon_1 v^2 \cos^2\beta \end{array}
\right)
\end{equation}

This mass matrix has one null eigenvalue, corresponding to the
Goldstone boson which ultimately lends a longitudinal component to the
Z boson. The other eigenstate is the pseudoscalar Higgs whose mass is
given by
\begin{equation}
\label{pseudo}
 M_A^2= \overline {m_3} ^2 (\tan \beta+\cot \beta) -2 \epsilon_1 v^2.
\end{equation}

The mass-matrix for the CP-even neutral Higgs bosons is now given by:
\begin{equation}
\label{higgs_matrix}
\mathcal{M}^2_{Re}=\left( \begin{array}{cc} (M_A^2 \sin^2\beta +M_Z^2
  \cos^2\beta)+2v^2(\epsilon_1 \sin^2 \beta + \epsilon_2 \cos^2
  \beta)& - (M_A^2+M_Z^2)\sin \beta \cos \beta+\epsilon_2 v^2 \sin
  \beta \cos \beta \\ -(M_A^2+M_Z^2)\sin \beta \cos \beta+\epsilon_2
  v^2 \sin \beta \cos \beta & (M_A^2 \cos^2\beta +M_Z^2
  \sin^2\beta)+2v^2(\epsilon_1 \cos^2 \beta + \epsilon_2 \sin^2
  \beta) \end{array} \right)
\end{equation}
Diagonalization of this mass-matrix will give the eigenvalues for the
neutral Higgs bosons:
\begin{equation}
\label{eigen}
 m^2_{H_0,h_0}=\frac{1}{2}[Tr \mathcal{M}^2_{Re} \pm \sqrt{(Tr
     \mathcal{M}^2_{Re})^2-4Det \mathcal{M}^2_{Re}}].
\end{equation}

We can calculate the lightest CP-even Higgs mass in the decoupling
limit from eqns. ~\ref{higgs_matrix} and ~\ref{eigen}.  We find that
\begin{equation}
\label{tree_corr}
 m^2_{h_0}=M_Z^2 \cos^2 2\beta + 2v^2 [2 \epsilon_1 \cos^2\beta
   \sin^2\beta + \epsilon_2
   (\cos^4\beta+\cos^2\beta\sin^2\beta+\sin^4\beta)]
\end{equation}

Thus at the tree level itself, the Higgs mass gets enhanced due to the
additional hard SUSY breaking terms. To this we must add the loop
corrections. In the absence of the dimensionless SUSY breaking terms,
the dominant correction to the lightest Higgs boson mass comes from
the contribution of the top quark and top squark loops. With the
assumption of a small mixing among the gauge eigenstates in the top
squark sector and with masses $m_{\tilde{t_1}}$, $m_{\tilde{t_2}}$
much greater than the top quark mass $m_t$, one finds a large positive
one-loop radiative correction to the Higgs mass~\cite{radiative}:

\begin{equation}
\label{soft_corr}
\Delta (m^2_{h_0}) \simeq \frac{3}{2 \pi^2} \frac{m_t^4}{v^2} \ln
(\frac{m_{\tilde{t_1}} m_{\tilde{t_2}}}{m_t^2}).
\end{equation}

Due to the presence of dimensionless hard couplings in the theory, in
addition to this there will be corrections quadratically sensitive to
the cut-off scale. They are generically given by:

\begin{equation}
 \delta m^2_{h_0} \sim \frac{\lambda_{hard}}{16 \pi^2} \Lambda^2,
\end{equation}

where $\lambda_{hard}$ is a generic dimensionless hard coupling giving
rise to quadratic one-loop corrections, and $\Lambda$ is the cut-off
of the theory.

In the case that we consider, we see from eqn.~\ref{V_Hard} that the
correction can be estimated to be

\begin{eqnarray}
\label{hard_corr}
 \delta m^2_{h_0} &\sim& \frac{\epsilon_2}{16 \pi^2} M^2 \nonumber
 \\ &=& \frac{1}{16 \pi^2} \frac{|F|^2}{M^4} M^2 \nonumber \\ &=&
 \frac{m_{soft}^2}{16 \pi^2}.
\end{eqnarray}

Note that, although this correction comes from a quadratically
divergent one-loop amplitude, it is bounded from above by the soft
SUSY-breaking mass squared, and is also independent of the cut-off
scale as long as the dimensionless couplings are of the given form (in
particular determined in terms of $F$ and $M$ and not arbitrary).

Adding up all the contributions from eqns.~\ref{tree_corr},
\ref{soft_corr} and \ref{hard_corr} in quadrature we find that the
Higgs mass for $\tan \beta \rightarrow 0$ can be as large as $155
\gev$ for a soft SUSY breaking scale $m_{soft} \sim 1 \tev$ and $M=
10\tev$. Thus the low SUSY breaking scale engineered by the extra
compact dimension(s) leads to a substantial enhancement of the Higgs
mass upper limit, so much so that the lightest neutral scalar in SUSY
may even be detected via the `gold-plated' $ZZ$ channel. It is also to
be noted that the dependence of the Higgs mass shift on $\tan \beta$
is rather weak, unless the cut-off scale $M$ is as low as $1
\tev$~\cite{Polonsky_1}.

In order to see the dependence of the lightest Higgs mass on the
cut-off scale $M$, for a fixed value of $M_A$, $\tan \beta$, and
$m_{soft}$, we diagonalize the CP-even scalar mass matrix in
eqn.~\ref{higgs_matrix} numerically for different values of $M$ and
present the results in Table~\ref{tab1}.  $F_T$ has also been varied
accordingly along with $M$ in order to keep $m_{soft}$ fixed at $1
\tev$. From Table~\ref{tab1} we can see that for lower values of $M$,
in the range of $1-5 \tev$, the enhancement in the lightest Higgs mass
is rather dramatic. Even the tree-level Higgs mass gets corrected by a
large amount. This is because the hard SUSY-breaking quartic couplings
in these cases become considerably large (for e.g., $\epsilon_1 \sim
\mathcal{O} (1)$ for $M \sim 1 \tev$) and comparable to, or more than,
the MSSM quartic coupling. In addition to the well known 1-loop
correction due the top-stop loops, there is also now an additional
1-loop contribution coming from the quadratically divergent graphs,
giving a further upward shift to the Higgs mass. This correction, as
noted before, is independent of the cut-off scale $M$, which is an
important feature of the form of the dimensionless couplings
involved. Thus, even if the cut-off scale is as high as $50 \tev$, we
obtain at least $\sim 20 \gev$ shift to the Higgs mass compared to the
1-loop prediction in MSSM. In addition, we can see from
eqns.~\ref{pseudo} and ~\ref{tree_corr} that, due to the dimensionless
SUSY breaking couplings, the scalar and pseudoscalar masses at the
tree level receive corrections of opposite sign. Thus, in our scenario
while the Higgs mass is predicted to increase compared to the MSSM
value, the pseudoscalar mass is expected to become lower.
\begin{table}[htb]
 \centering
%\begin{ruledtabular}
 \begin{tabular}{|c| c| c|}
\hline \hline
M & $m^{tree}_{h_0}$&$m^{tree+1-loop}_{h_0}$\\
 ($\tev$)   & ($\gev$)             &($\gev$)             \\
\hline
1 & 361 &382 \\

5 & 115 &170 \\

10 & 95 &156 \\

50 & 83 &150 \\
\hline \hline
 \end{tabular}

%\end{ruledtabular}
\caption{\label{tab1} Change in mass of the lightest Higgs boson with
  variation in the cut-off scale M. The masses have been calculated
  with $M_A= 200 \gev$, $\tan \beta =5$ and $m_{soft} = 1 \tev$. $F_T$
  has been varied concomitantly with $M$.}
\end{table}
%%%%%%%%%%%%%%%%%%%%%%%%%%%%%%%%%%%%%%%%%%%%%%%%%%%%%%%%%%%%%%%%%%%

%%%%%%%%%%%%%%%%%%%%%%%%%%%%%%%%%%%%%%%%%%%%%%%%%%%%%%%%%%%%%%%%%%%

Finally, let us also note that the model considered by us here does
not involve any additional matter or gauge fields. In that sense, we
still work in a ``minimal'' framework, where the SUSY-breaking terms
are suppressed by a lower scale and therefore the effect of hard
SUSY-breaking operators become important. And we observe that it is
possible to obtain a substantial enhancement in the lightest Higgs
mass, thereby relaxing the stringent upper bound found in MSSM. In
addition, if the cut-off scale is rather low, even the tree-level
Higgs mass gets significantly shifted, thus alleviating the tension
between the LEP limit and the MSSM bound on the Higgs
mass~\cite{casas}.  In other approaches to solving this so-called
fine-tuning problem within the MSSM, there have been studies where the
MSSM is extended, for example, by an additional $U(1)$ gauge group,
accompanied by a singlet scalar field coupling to the two Higgs
doublets by a superpotential term~\cite{U_1}. Although the upper bound
on the tree-level lightest CP-even Higgs doublet mass is raised in
these models, too, the actual mass eigenvalues are generally smaller
because of mixing with the singlets. The tension with the LEP-bound is
still avoided as the LEP limits themselves change due to modifications
in the relevant couplings~\cite{U_1}. Therefore, we note that the
Higgs mass eigenvalues predicted in these extensions of the MSSM often
have much lower magnitude than those obtained in our scenario. This
can be used as a discriminating feature of our scenario with these
gauge and singlet extensions of the MSSM. Besides, the scheme proposed
by us is offered as an explanation of a situation where the signals of
SUSY (say, in the form of large missing transverse momentum and
jets/leptons) are found at the LHC, but the lightest neutral Higgs
mass reconstructed turns out to be considerably higher than what is
normally expected within the MSSM.

\section{Summary and Conclusion}
\label{summary}

We have considered a scenario where the soft SUSY-breaking terms are
generated through higher-dimensional operators suppressed by energy
scales a little above a TeV.  At the same time, possible hard SUSY
breaking terms have been retained, which can contribute to the Higgs
mass(es) at tree-level as well as through loop effects.  This is a
generic situation that can be expected if new physics is always
staring one in the face as one goes continuously upwards in energy,
rather than there being some `desert' above the TeV order. We
demonstrate the viability of our proposal in an illustrative model
based on extra compact spacelike dimensions. Here one indeed obtains
such a scale, corresponding to the stable vev of the scalar component
of the modular fields connected with the extra dimensions.  A scalar
potential has been explicitly constructed where both the SUSY breaking
F-term vev ($F_T$) and the modular field vev ($\langle T \rangle$) lie
in the range of a few TeV's, and give rise to TeV-scale soft breaking
parameters. The vev of $T$ sets the scale of suppression of
higher-dimensional operators at a scale below $\langle T\rangle$ when
$T$ is integrated out, and thus generates the predominant
SUSY-breaking effects.

We show that, in such a setting, the lightest even-parity neutral
Higgs mass can receive two types of additional contributions: one at
the tree-level due to the hard SUSY-breaking term(s), and the other at
the loop level, dictated by the new cut-off scale of $10 \tev$ or
so. Our numerical estimate shows that, over the usual,
phenomenologically allowed region of the MSSM parameter space, this
leads to considerable upward revision of the upper limit of the
lightest Higgs mass. While for $\langle T \rangle \simeq 1 \tev$, the
corrected mass limit can be as large as about $380 \gev$, more modest,
but significant, revisions upto $150 - 170 \gev$ are quite possible
for cut-off scales upto $10 \tev$.  Thus, if signals of SUSY, perhaps
in the form of large missing energy and jets/leptons, reveal
themselves at the Large Hadron Collider (LHC), and at the same time
one fails to find the Higgs boson with mass below 130 GeV or so, the
situation is, after all, not irreconcilable. One should then look
seriously at the possibility of the clue to SUSY breaking lying a
little above the reach of the LHC, with possible indirect
manifestations in other experimental results. The fact that the fields
connected with the fluctuating radii of extra dimensions can be
responsible for such large shift in the Higgs mass injects an added
degree of richness to such a scenario.

\section*{Acknowledgement}
This work was partially supported by funding available from the
Department of Atomic Energy, Government of India, for the Regional
Centre for Accelerator-based Particle Physics (RECAPP), Harish-Chandra
Research Institute. We thank Rajesh Gopakumar and Andreas Nyffeler for
useful discussions.  SSG acknowledges the hospitality of RECAPP where
this project was initiated. S.M. and B.M. also thank the Indian
Association for the Cultivation of Science, Kolkata, for hospitality
while work was in progress.

%%%%%%%%%%%%%%%%%%%%%%%%%%%%%%%%%%%%%%%%%%%%%%%%%%%%%%%%%%%%%%%%%%%%%%%%%%%%%%%%%%%%%%%%

\end{document}